# Inferring solar-wind plasma structures from sparse probe trajectories using recurrent reduced-order learning

Maryam Reza[*], Farbod Faraji[1**]

[*] Independent Researcher, London, United Kingdom

[**] Department of Computing, Imperial College London, London, SW7 2RH, United Kingdom

**Abstract**: In space plasma studies, spacecraft measurements often provide time histories of the solar-wind plasma. However, many heliospheric plasma processes are organized over spatial scales that cannot be directly resolved by limited local sampling. This creates a persistent challenge: how to use limited probe measurements to recover the spatial plasma distributions needed to interpret evolving solar-wind structures. In this work, we present a recurrent reduced-order learning framework to address this challenge. The method is demonstrated using WSA-ENLIL solar-wind simulation data to reconstruct two-dimensional meridional and equatorial fields from a small number of virtual probes, with radial velocity and plasma density considered as target quantities on both planes. From sparse temporal probe signals as inputs, the model recovers the dominant radial and latitudinal variations in the meridional plane and the spiral-shaped organization of the equatorial solar wind. It is also able to reconstruct spatial distributions of dynamically coupled plasma fields not directly sensed. Sensitivity studies are performed to assess the dependence of reconstruction accuracy on key parameters of the machine-learning framework: modal rank, number of probes, and input-history length. The outcomes underline the methodology's promise as a practical route for extracting spatial plasma-state information from spacecraft measurements in support of studies on the underlying physics of space and solar-wind plasmas.

## 1. Introduction

**Heliospheric plasma and significance of space weather**. Understanding the multi-scale structure and dynamics of heliospheric plasma remains a central question in space physics and space weather research. The heliosphere is a plasma environment extending from the solar corona to the outer boundaries of the solar system, through which solar-wind structures propagate toward Earth and other planetary systems. Through this propagation, variations in solar-wind plasma properties drive a broad range of phenomena, including coronal mass ejections (CMEs), co-rotating interaction regions (CIRs), shocks, turbulence, geomagnetic storms, and magnetospheric disturbances. These processes affect satellite operations, navigation and communication systems, and astronaut safety. Accurate characterization and prediction of heliospheric plasma states are thus essential for both scientific understanding and operational space-weather forecasting.

**Sparse observations and the reconstruction challenge**. Unlike laboratory plasmas, where diagnostics can often be deployed throughout a controlled domain, heliospheric plasma is sampled by a limited number of spacecraft and remote-sensing instruments. The distributed plasma state must therefore be inferred from sparse in-situ measurements, remote observations, and physics-based simulations. This makes full-state reconstruction a key challenge in space plasma physics: the system is global, three-dimensional, and structured by coronal magnetic topology, solar rotation, stream interaction regions, CMEs, shocks, turbulence, and kinetic-scale processes [1]-[4], while the available measurements are typically local, limited, and non-uniformly distributed. This disparity between global plasma structure and local measurement access motivates methods that can extract distributed-state information from sparse time-resolved observations.

**From spacecraft time series to distributed plasma states**. Spacecraft missions such as ACE, Wind, Ulysses, STEREO, Parker Solar Probe, Solar Orbiter, and planetary missions have provided transformative measurements of heliospheric plasma. These observations provide detailed local time series of solar-wind conditions along specific spacecraft trajectories, but converting such measurements into distributed plasma states requires additional assumptions. Single-spacecraft time series are often interpreted using Taylor's hypothesis [5]-[7], ballistic propagation [8], co-rotation mapping [9], or model-assisted extrapolation [10]. These approaches can be effective in slowly evolving or weakly interacting solar-wind streams [8]-[10][, but their accuracy can degrade in accelerating flows, interacting streams, CME sheaths, shocks, or strongly evolving turbulent regions. Remote heliospheric imaging and solar observations provide complementary information about large-scale structures, solar sources, and propagation context. However, line-of-sight integration, projection effects, uncertain magnetic boundary conditions, and limited cadence make it difficult to uniquely infer the evolving plasma distribution throughout the heliosphere from these observations alone [11]-[13]. Thus, heliospheric plasma reconstruction from sparse probes is not simply a spatial interpolation problem, but an underdetermined dynamical inverse problem constrained by plasma physics, sparse sensing, and the statistical structure of the solar wind.

[1] **Corresponding author**: f.faraji20@imperial.ac.uk

**Physics-based heliospheric modeling**. Physics-based modeling remains the primary route for obtaining globally resolved heliospheric plasma states. Among the most widely used operational heliospheric models is the Wang-Sheeley-Arge (WSA) model coupled with the ENLIL magnetohydrodynamic (MHD) solver. The WSA model uses photospheric magnetic-field observations to estimate solar-wind properties near the corona [14][15], while ENLIL propagates the estimated ambient solar-wind conditions throughout the heliosphere using three-dimensional MHD simulations [16]-[18]. WSA-ENLIL is widely used by NOAA's[1] Space Weather Prediction Center and numerous international forecasting centers. The Magnetohydrodynamics Around a Sphere (MAS) model provides another widely used physics-based description of the corona and inner heliosphere, with applications to coronal magnetic-field structure, solar-wind acceleration, and CME propagation [19]. The SUSANOO-SW model similarly solves the global heliospheric MHD evolution and has been used for space-weather prediction and interplanetary CME propagation studies [20]. In addition, the Space Weather Modeling Framework (SWMF) provides a broader modular Sun-to-Earth modeling environment, coupling solar-coronal, heliospheric, magnetospheric, ionospheric, and upper-atmosphere components, many of which are based on the adaptive-grid BATS-R-US MHD solver [21]-[24].

**The gap between global simulations and sparse measurements**. Despite these capabilities, significant challenges remain. Global heliospheric simulations are computationally expensive, particularly when ensemble forecasting, uncertainty quantification, parameter studies, or long-duration simulations are required [10][25]. Their predictive accuracy also depends strongly on uncertain solar boundary conditions, incomplete physical descriptions, reduced plasma closures, and unresolved multiscale processes. More complete descriptions, including hybrid, Vlasov, and particle-in-cell methods, can represent additional kinetic physics, but become computationally prohibitive at global heliospheric scales [26]. Consistent with these limitations, model validation studies [27]-[29] demonstrate that significant discrepancies can remain between predicted and observed solar-wind conditions, highlighting the need for improved state-estimation methodologies and stronger integration between observations and simulations. Consequently, there remains a methodological gap between sparse spacecraft measurements and expensive high-dimensional simulations: the need for approaches that retain the global structure learned from physics-based models while enabling rapid reconstruction from limited observational information.

**Reduced-order and data-driven modeling**. Addressing the gap between global simulations and sparse measurements requires methods that can represent the dominant structure of high-dimensional plasma fields in a compact and computationally efficient form. Reduced-order modelling and data-driven methods provide one route to this goal by exploiting coherent spatiotemporal organization in complex dynamical systems [30]. Techniques such as dynamic mode decomposition (DMD), Koopman-based analysis, sparse identification of nonlinear dynamics (SINDy), and deep neural networks have been used to extract coherent structures, learn reduced representations, approximate high-dimensional dynamics, and develop fast surrogate models at reduced computational cost [31]-[35]. In plasma physics, related methods have been applied to the discovery of approximate governing dynamics, modal analysis, reduced-order modeling, and surrogate prediction [30][36]-[39]. These approaches exploit the fact that high-dimensional plasma data often contain lower-dimensional coherent structure, allowing much of the dominant variability to be represented using a limited number of modes, coordinates, or latent variables. Nevertheless, reduced-order and data-driven models do not by themselves solve the sparse-observation problem. Their long-horizon predictive fidelity can be limited by error accumulation and phase drift, and, for reconstruction from spacecraft data, the key challenge is not only to define a compact dynamical representation, but to infer its evolving state from very limited measurements.

[1] National Oceanic and Atmospheric Administration

**Compressed sensing under spacecraft constraints**. Data-assimilation and compressed-sensing techniques address this inference problem more directly by combining sparse observations with physical or statistical representations of the underlying system state. Compressed sensing offers a conceptually attractive framework for reconstructing high-dimensional signals from limited observations when the state is sparse or compressible in a suitable basis [40]-[42]. In this setting, the central idea is that many high-dimensional physical systems can be represented in terms of low-dimensional global modes and can therefore, in principle, be reconstructed from a limited number of measurements. Related approaches include variational data assimilation, ensemble Kalman filtering, Gaussian-process methods, Bayesian inference, and sparse-sensing techniques that combine Proper Orthogonal Decomposition (POD) bases with optimized sensor placement or gappy reconstruction methods [43]-[47]. These methods seek to combine observations with physics-based models to produce improved estimates of the underlying system state. However, their practical application often requires a relatively large number of sensors, carefully designed measurement operators, or optimized sensor-placement strategies. These requirements become particularly restrictive in space-plasma applications, where sensor locations are dictated by mission design and spacecraft trajectories rather than by optimal experimental design. In addition, the most informative regions may be inaccessible, unmeasured, or sampled only intermittently. Hence, a practical heliospheric reconstruction method must operate with very few sensors, tolerate non-optimal probe locations, and exploit the temporal richness of spacecraft time series rather than relying on large spatial sample sets.

**Machine learning beyond scalar space-weather prediction**. Machine learning (ML) has become increasingly important in space-weather and heliophysics research, with applications including solar-flare prediction, CME arrival-time forecasting, geomagnetic-index prediction, radiation-belt modeling, and solar-wind forecasting [48]-[50]. These applications demonstrate the value of data-driven learning for extracting predictive information from complex heliophysical data. However, most existing studies focus on event classification or the prediction of scalar quantities at specific locations, such as near-Earth solar-wind parameters. The reconstruction problem considered here is different: the goal is to infer a high-dimensional spatial plasma distribution from minimal probe measurements. This requires a model that can combine the compact global representation used in reduced-order modelling with the sparse-input philosophy of compressed sensing, while using temporal probe histories (time trajectories) to compensate for the lack of dense spatial measurements.

**Paper contribution**. This paper demonstrates a recurrent reduced-order learning framework for inferring spatial distributions of solar-wind plasma structures from a minimal number of sparse, time-resolved probe measurements. The central contribution is to assess the utility of this reconstruction methodology for heliospheric plasmas, where spacecraft routinely provide finite-duration time series while simultaneous spatial coverage remains intrinsically limited and sparse.

The introduced model combines temporal sequence learning, nonlinear state reconstruction, and a compressed representation of global plasma fields. The architecture consists of a recurrent temporal encoder, based on long short-term memory (LSTM) networks, coupled to a shallow nonlinear decoder [51]. The recurrent component learns the temporal information contained in sparse probe histories, while the decoder maps the resulting latent representation to a reduced POD representation of the plasma state. The full-domain field is then reconstructed by projection onto precomputed spatial POD modes.

The central advantage of this architecture is that it uses the temporal richness of probe histories to compensate for limited spatial sampling, thus reducing dependence on dense or optimally placed sensor arrays. Previous applications of the architecture to low-temperature laboratory and industrial plasmas, including those relevant to spacecraft electric propulsion, have shown that full plasma states can be reconstructed from as few as three randomly placed local probes measuring a single plasma property [52][53]. Importantly, the reconstructed state can include dynamically coupled quantities that are not directly measured [52].

The present paper demonstrates this ML-based sparse-sensing and inference methodology using WSA–ENLIL solar-wind prediction data as a controlled heliospheric test case. Virtual probe trajectories are extracted from simulated fields and used to reconstruct meridional and equatorial plasma distributions. The method is further assessed through sensitivity studies on the retained POD rank, the number of input probes, and the input lag length, thereby quantifying how reconstruction accuracy depends on reduced-order representation, spatial sparsity, and temporal information content.

## 2. Methodology

### 2.1. Problem formulation

The objective of this work is to reconstruct a high-dimensional heliospheric plasma state from a small number of time-series probe measurements. This problem can be formulated as follows: let the full plasma state at time $t_i$ be denoted by state vector $\mathbf{x}_i$, where

$$\mathbf{x}_i \in \mathbb{R}^n, \quad \text{(Eq. 1)}$$

and $n$ is the number of spatial degrees of freedom after discretizing the plasma field. If multiple plasma quantities are reconstructed simultaneously, the full state can be written as a concatenated multi-field vector,

$$\mathbf{x}_i = \left[\mathbf{x}_i^{(1)}, \mathbf{x}_i^{(2)}, \dots, \mathbf{x}_i^{(P)}\right], \quad \text{(Eq. 2)}$$

where $P$ is the number of plasma variables, such as density, velocity, pressure, or magnetic-field components.

Only a small number of measurements are assumed to be available. These sparse measurements are represented as

$$\mathbf{y}_i = \mathbf{C}\mathbf{x}_i, \quad \text{(Eq. 3)}$$

where $\mathbf{y}_i \in \mathbb{R}^m$, $m \ll n$, and $\mathbf{C}$ is the measurement operator. For pointwise virtual probes, $\mathbf{C}$ selects the grid locations corresponding to the sensor positions. More general measurement operators may also be used, including local averages, integrated quantities, or other reduced observational signals. Rather than using only the instantaneous measurement $\mathbf{y}_i$, the model uses a finite time history of probe measurements. For a time window of length $K$, the input sequence is expressed as

$$\mathcal{Y}_i = [\mathbf{y}_{i-K+1}, \mathbf{y}_{i-K+2}, \dots, \mathbf{y}_i]. \quad \text{(Eq. 4)}$$

Thus, the reconstruction problem is to learn an optimal nonlinear mapping from sparse time-history measurements to the full plasma state, $\mathcal{F}: \mathbb{R}^{K\times m} \to \mathbb{R}^n$, such that

$$\hat{\mathbf{x}}_i = \mathcal{F}(\mathcal{Y}_i). \quad \text{(Eq. 5)}$$

$\hat{\mathbf{x}}_i$ approximates the true plasma state $\mathbf{x}_i$, and $\mathcal{F}$ is learned to minimize the mean-squared discrepancy between the reconstructed and true plasma states

$$\mathcal{F} = argmin \frac{1}{N_s}\sum_{i=1}^{N_s} \|\mathbf{x}_i - \hat{\mathbf{x}}_i\|_2^2, \quad \text{(Eq. 6)}$$

where $N_s$ is the number of available data samples. This formulation uses temporal information from a small number of probes to compensate for the absence of dense spatial measurements.

### 2.2. Low-rank representation of the plasma state

Learning directly the mapping from sparse probe time series to full plasma state can be computationally expensive when the spatial dimension is large. In addition, defining the reconstruction objective directly in physical space through a node-by-node mean-squared error can bias the optimization toward the largest-amplitude and spatially averaged features of the data. Since large-scale structures usually dominate the total variance of the plasma field, smaller-amplitude structures, which are still dynamically relevant, may contribute less to the loss function and can hence be missed from the learning.

To mitigate this issue, the plasma fields are first projected onto a reduced modal basis obtained from singular value decomposition (SVD), equivalently POD, for snapshot data. The retained number of bases/ modes is chosen to represent sufficient variations and coherent patterns that are of interest and relevant to the reconstruction. The learning problem is then formulated in this reduced representation.

For the $p$-th plasma quantity, the snapshot matrix is assembled as

$$\mathbf{X}^{(p)} = \left[\mathbf{x}_1^{(p)}, \mathbf{x}_2^{(p)}, \dots, \mathbf{x}_N^{(p)}\right], \quad \text{(Eq. 7)}$$

where,

$$\mathbf{X}^{(p)} \in \mathbb{R}^{n_p \times N}. \quad \text{(Eq. 8)}$$

In Eq. 8, $n_p$ is the spatial dimension of the $p$-th field, and $N$ is the number of temporal snapshots. The SVD of the matrix $\mathbf{X}^{(p)}$ is

$$\mathbf{X}^{(p)} = \mathbf{U}^{(p)}\mathbf{\Sigma}^{(p)}\mathbf{V}^{(p)T}. \quad \text{(Eq. 9)}$$

A rank-$r$ approximation is then obtained by retaining only the leading $r$ modes, as per Eq. 10

$$\mathbf{X}_{\mathbf{r}}^{(p)} \approx \mathbf{U}_r^{(p)}\mathbf{\Sigma}_r^{(p)}\mathbf{V}_r^{(p)T}. \quad \text{(Eq. 10)}$$

In Eq. 10, $\mathbf{U}_r^{(p)}$contains the dominant spatial modes, $\mathbf{\Sigma}_r^{(p)}$contains the leading singular values, and $\mathbf{V}_r^{(p)}$contains the temporal coefficients of the corresponding modes.

Once data are projected on the rank-$r$ SVD (POD) modes, instead of training the ML model to reconstruct the full spatial field directly, the model is tasked to reconstruct a reduced coefficient vector, $\mathbf{a}_i^{(p)} \in \mathbb{R}^r$, which is a column of $\mathbf{V}_r^{(p)T}$ at each time instance. The reconstructed $p$-th plasma field is then obtained by projecting the predicted reduced coefficients $\hat{\mathbf{a}}_i^{(p)}$ back to physical space as in Eq. 11

$$\hat{\mathbf{x}}_i^{(p)} = \mathbf{U}_r^{(p)}\mathbf{\Sigma}_r^{(p)}\hat{\mathbf{a}}_i^{(p)}. \quad \text{(Eq. 11)}$$

**2.3. The recurrent reduced-order ML model architecture**

The ML model consists of two main components: a recurrent LSTM network as the temporal encoder and a shallow feed-forward fully connected network as a nonlinear decoder. The recurrent encoder learns the temporal information contained in the time histories of sparse probe signals, while the decoder maps this learned temporal representation in the latent space of the encoder to the reduced plasma-state represented by columns of $\mathbf{V}_r^T$ from the POD, which will then be converted back to the physical space using Eq. 10.

The input to the recurrent encoder is the time-lagged measurement sequence $\mathcal{Y}_i$ (Eq. 4). The encoder maps this sequence to a latent hidden representation, $\mathbf{h}_i = \mathcal{G}(\mathcal{Y}_i; \boldsymbol{\theta}_r)$, where $\mathcal{G}$ denotes the recurrent encoder and $\boldsymbol{\theta}_r$ represents its trainable parameters. At each time step, the recurrent network updates its hidden and memory states using the current measurement and the previous internal states. This update can be written compactly as

$$(\mathbf{h}_i, \mathbf{c}_i) = \mathcal{L}(\mathbf{y}_i, \mathbf{h}_{i-1}, \mathbf{c}_{i-1}; \boldsymbol{\theta}_r), \quad \text{(Eq. 12)}$$

where $\mathbf{h}_i$ is the hidden state, $\mathbf{c}_i$ is the memory state, and $\mathcal{L}$ represents the recurrent update rule. The final hidden representation (state) encodes the temporal information contained in the sparse measurement history.

The feed-forward nonlinear decoder network subsequently maps that hidden representation to the reduced plasma-state coefficients, $\hat{\mathbf{a}}_i = \mathcal{D}(\mathbf{h}_i; \boldsymbol{\theta}_d)$, in which $\mathcal{D}$ is the decoder, $\boldsymbol{\theta}_d$ denotes its trainable parameters, and $\hat{\mathbf{a}}_i$ is the predicted reduced-rank plasma state vector. For a decoder with $B$ layers, the mapping can be written schematically as

$$\mathcal{D}(\mathbf{h}) = \phi_B(\mathbf{W}_B\phi_{B-1}(\dots\phi_1(\mathbf{W}_1\mathbf{h}+\mathbf{b}_1)\dots)+\mathbf{b}_B). \quad \text{(Eq. 13)}$$

$\mathbf{W}_j$ and $\mathbf{b}_j$ are the trainable weights and biases of layer $j$, and $\phi_j$s are the nonlinear activation functions. The decoder is intentionally shallow to keep the number of trainable parameters moderate and to reduce the tendency to overfit, especially when the available training data are limited.

The complete ML model can therefore be expressed as

$$\hat{\mathbf{a}}_i = \mathcal{D}(\mathcal{G}(\mathcal{Y}_i; \boldsymbol{\theta}_r); \boldsymbol{\theta}_d). \quad \text{(Eq. 14)}$$

After the reduced coefficients are predicted, the full plasma field is reconstructed using the retained spatial basis

$$\hat{\mathbf{x}}_i = \mathbf{\Phi}_r \hat{\mathbf{a}}_i, \qquad \text{(Eq. 15)}$$

in which $\mathbf{\Phi}_r$ represents the retained spatial basis, including the corresponding singular-value scaling ($\mathbf{\Phi}_r = \mathbf{U}_r \mathbf{\Sigma}_r$). As mentioned earlier, in the present methodology workflow, the spatial modes are obtained from SVD of the training data.

For multiple plasma variables, separate reduced bases may be computed for each field. The model output can then be written as a concatenation of the predicted reduced coefficients for all target variables (Eq. 16)

$$\hat{\mathbf{a}}_i = \left[\hat{\mathbf{a}}_i^{(1)}, \hat{\mathbf{a}}_i^{(2)}, \ldots, \hat{\mathbf{a}}_i^{(P)}\right]. \qquad \text{(Eq. 16)}$$

Each coefficient vector is projected through its corresponding spatial basis to reconstruct the associated physical plasma field. The schematic of the ML architecture and the related workflow described in this section are presented in Figure 1.

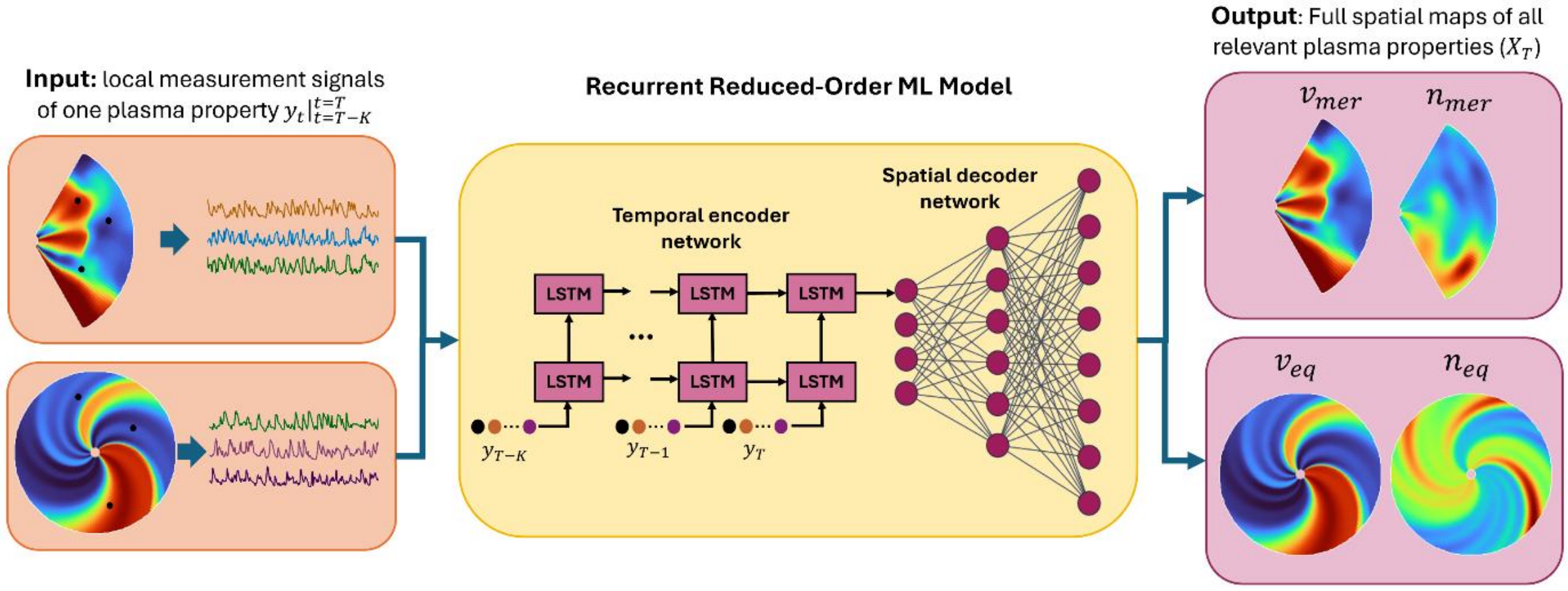

*Figure 1: Schematic visualization of the ML architecture; black dots on the sample plasma property maps on the left-hand side (input) represent the locations of virtual probes.*

The model is trained using pairs of sparse measurement histories and corresponding reduced plasma-state coefficients. The training dataset is written as $\{\mathcal{Y}_i, \mathbf{a}_i\}_{i=1}^{N_t}$, with $N_t$ being the number of training samples, $\mathcal{Y}_i$ the sparse measurement history, and $\mathbf{a}_i$ the target reduced coefficient vector corresponding to the SVD/POD representation of the ground-truth plasma field. The trainable parameters of the recurrent encoder and decoder are collected as $\boldsymbol{\theta} = \{\boldsymbol{\theta}_r, \boldsymbol{\theta}_d\}$.

The model parameters are obtained by minimizing the mean-squared error between the predicted and target reduced coefficients as per Eq. 17

$$\boldsymbol{\theta}^* = \underset{\boldsymbol{\theta}}{\arg min} \frac{1}{N_t} \sum_{i=1}^{N_t} \|\mathbf{a}_i - \hat{\mathbf{a}}_i\|_2^2. \qquad \text{(Eq. 17)}$$

Equivalently, the reconstruction error can be evaluated in the physical space after projection through the retained spatial modes, i.e.,

$$\epsilon = \|\mathbf{x}_i - \hat{\mathbf{x}}_i\|_2^2. \qquad \text{(Eq. 18)}$$

## 3. Application to heliospheric plasma and solar-wind structures

For the demonstrations of the ML model that follows, the sparse probe measurements are generated synthetically from the WSA-ENLIL simulation fields. Specifically, virtual sensors are placed at selected locations in the computational domain, and their time histories are extracted from the simulated plasma variables. This provides a controlled setting in which the true full-field solution is known and can be used to train and evaluate the reconstruction model. In practice, these synthetic sensor histories would be replaced by in-situ time-series measurements from spacecraft or plasma probes, while using the same reconstruction framework to infer the corresponding distributed heliospheric plasma state.

### 3.1. Dataset description and preprocessing

The model demonstration is performed using background solar-wind simulation data from the NOAA/NCEI WSA-ENLIL Solar Wind Prediction archive [54]. WSA-ENLIL provides three-dimensional (3D) heliospheric MHD simulations; in this work, we use the archived two-dimensional (2D) plasma-field distributions on the meridional and equatorial planes.

To curate the dataset, weekly WSA-ENLIL background runs were downloaded over the period of 23 May 2023 to 23 May 2026, using a seven-day interval between files to reduce overlap between consecutive simulation outputs. Four plasma-field datasets were derived from the downloaded runs: radial velocity on the meridional plane, $v_{mer}$; radial velocity on the equatorial plane, $v_{eq}$; plasma density on the meridional plane, $n_{mer}$; and plasma density on the equatorial plane, $n_{eq}$.

The meridional and equatorial fields were defined on $60 \times 512$ and $180 \times 512$ grids, respectively. Each snapshot was thus flattened before applying SVD, giving spatial dimensions of $n = 30{,}720$ for the meridional fields and $n = 92{,}160$ for the equatorial fields.

The dataset was organized as a sequence of WSA-ENLIL simulation runs, containing total of 25,350 temporal snapshots for each plasma field, corresponding to 150 WSA-ENLIL runs with 169 frames per run. Time-lagged samples were constructed within individual runs only, so that no input sequence crossed the boundary between two independent simulation files. A lag length of $K = 50$ was used; therefore, each 169-frame run produced 120 valid time-history windows. This resulted in 18,000 valid input-output samples in total. The dataset was split by simulation run rather than by individual snapshot. The first 105 runs, corresponding to 70% of the available runs, were used for training, while the remaining 45 runs were reserved for testing. This produced 12,600 training samples and 5,400 test samples.

Sparse input measurements were generated from the velocity fields. For meridional-plane reconstructions, virtual-probe signals were extracted from $v_{mer}$ and used to reconstruct both $v_{mer}$ and $n_{mer}$. Similarly, for equatorial-plane reconstructions, input signals were extracted from $v_{\mathrm{eq}}$ and used to reconstruct both $v_{eq}$ and $n_{eq}$. In this way, we are demonstrating that sparse time histories of one measured plasma variable can reconstruct both the measured field itself and a dynamically related field that had not been directly observed. Comparable reconstruction performance was observed when the corresponding plasma-density probe signals were used as inputs to reconstruct the full density and velocity fields, indicating that either dynamically coupled variable can provide sufficient temporal information for full-state reconstruction.

For each plane, meridional or equatorial, five virtual probes were selected randomly within the spatial domain. Each probe signal was defined as the average over a $6 \times 6$ grid-point block centered on the probe location, rather than as a single grid-point value, to reduce sensitivity to grid-scale numerical fluctuations. This local averaging was used to construct numerically robust synthetic probe measurements from the gridded WSA-ENLIL data and can be interpreted as a finite-resolution virtual measurement, loosely analogous to the finite sampling volume associated with real space-borne plasma measurements.

### 3.2. Model setup and implementation parameters

Four separate ML-model instances with similar settings and architecture were trained, one for each target field: $v_{mer}$, $v_{eq}$, $n_{mer}$, and $n_{eq}$. For each target field (property), a randomized SVD [55] was computed on the training data to obtain the reduced modal representation. Randomized SVD was used instead of full (standard deterministic) SVD for computational efficiency, given the high dimensionality of the dataset. 30 modes ($r$ = 30) were retained for each field/property. The model was trained to predict the retained reduced modal coefficients rather than the full spatial field directly. After prediction, the coefficients were projected back to physical space using the retained POD basis ($\mathbf{\Phi}_r$), as was described in Section 2.3.

Each model instance used an LSTM temporal encoder with two recurrent layers and a hidden dimension of 32, followed by a fully connected decoder consisting of two hidden layers with widths 100 and 150 and a final linear output layer. Rectified linear unit (ReLU) activation functions were used after the hidden layers of the fully connected decoder. The size of the final layer was equal to the number of retained modes for the corresponding target field, which is 30 in this case. A dropout rate of 0.1 was used in the decoder during training. The models were trained for 300 epochs using a batch size of 64 and a learning rate of $10^{-3}$.

## 4. Results

The trained models were evaluated on unseen time intervals (test data) by reconstructing the full spatial plasma property maps from sparse probe time series alone. The reconstructed fields were compared with both the ground truth rank-truncated SVD fields and the original full-rank WSA-ENLIL snapshots.

Figure 2 shows the singular-value spectra and SVD reconstruction error for the four plasma-field datasets. The error of the rank-$r$ SVD approximation was quantified using the relative Frobenius-norm error between the full snapshot matrix $\mathbf{X}^{(p)}$ and its rank-truncated reconstruction $\mathbf{X}_r^{(p)}$ as per the following relation

$$\epsilon_{\mathrm{SVD}}(r) = \frac{\|\mathbf{X}^{(p)} - \mathbf{X}_r^{(p)}\|_F}{\|\mathbf{X}^{(p)}\|_F}, \quad \text{(Eq. 19)}$$

in which, $\|\cdot\|_F$ denotes the Frobenius norm.

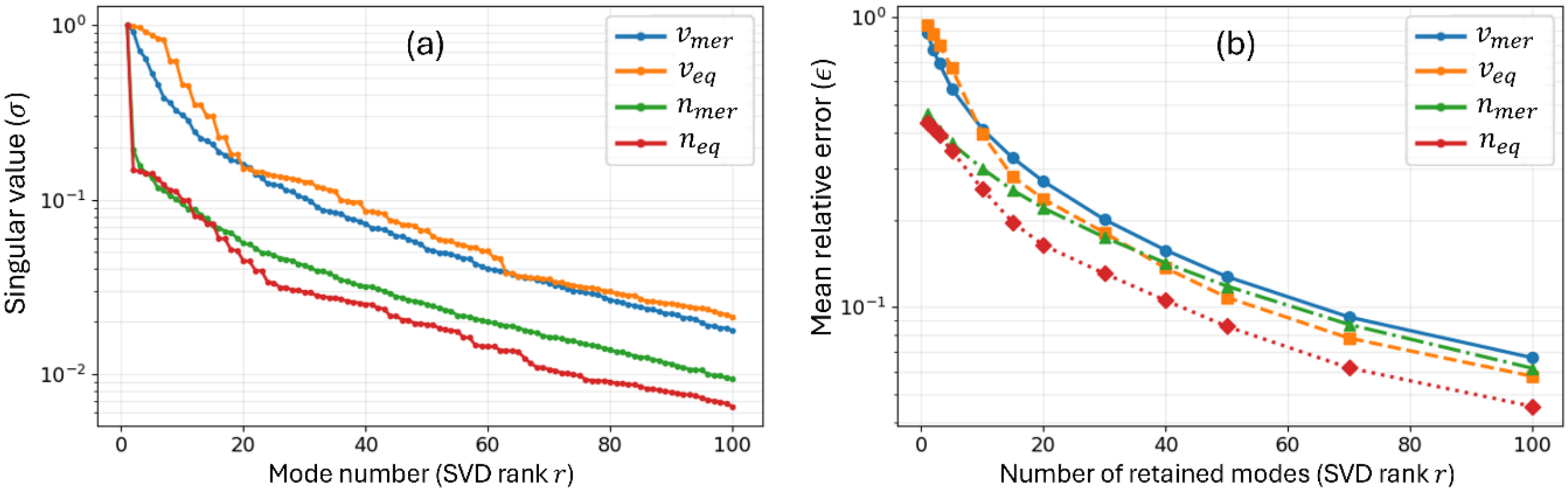

*Figure 2: Low-rank structure of the WSA-ENLIL plasma fields obtained from SVD; (a) singular-value spectra and (b) SVD reconstruction error (obtained from Eq. 19) vs the number of retained modes for $v_{mer}$, $v_{eq}$, $n_{mer}$, and $n_{eq}$.*

From Figure 2, it is seen that at low ranks, the SVD error decreases rapidly and then more gradually as higher modes are added. By $r = 30$, the relative reconstruction error is reduced to approximately 0.15–0.20 for all four fields, while increasing the rank to $r = 100$ lowers the error further to about 0.04–0.06. For the results presented in this section, the 30 leading modes are retained, providing a compact reduced representation while capturing most of the dominant large-scale heliospheric structure. Higher modes mainly refine smaller-amplitude features. The selected rank is also informed by the sensitivity analysis in Section 5.1, which shows that increasing the retained rank beyond a certain limit does not significantly change the ML-model reconstruction accuracy.

Figure 3 to Figure 6 compare the ML-model inference of full spatial distributions of $v_{mer}$, $n_{mer}$, $v_{eq}$ and $n_{eq}$ against the ground-truth rank-truncated SVD fields and the full-rank WSA-ENLIL fields for randomly selected test snapshots. Across all four quantities, the inferred fields closely follow the rank-truncated SVD targets, indicating that the model has learned the mapping from sparse probe histories to the retained modal representation. The agreement with the full-rank WSA-ENLIL fields is strongest for the dominant larger-scale structures, while discrepancies are mainly associated with finer features that are either weakly represented in the retained rank or more difficult to infer from sparse measurements. Figure 3 and Figure 4 show that the model has recovered the main radial and latitudinal variations in the meridional field reconstructions, while Figure 5 and Figure 6 show that the model captures the spiral-like organization of the solar-wind structure in the equatorial plane. Importantly, the density fields are reconstructed from velocity-based probe inputs, demonstrating that the learned latent representation contains information about dynamically coupled plasma quantities, not only the directly sampled variable.

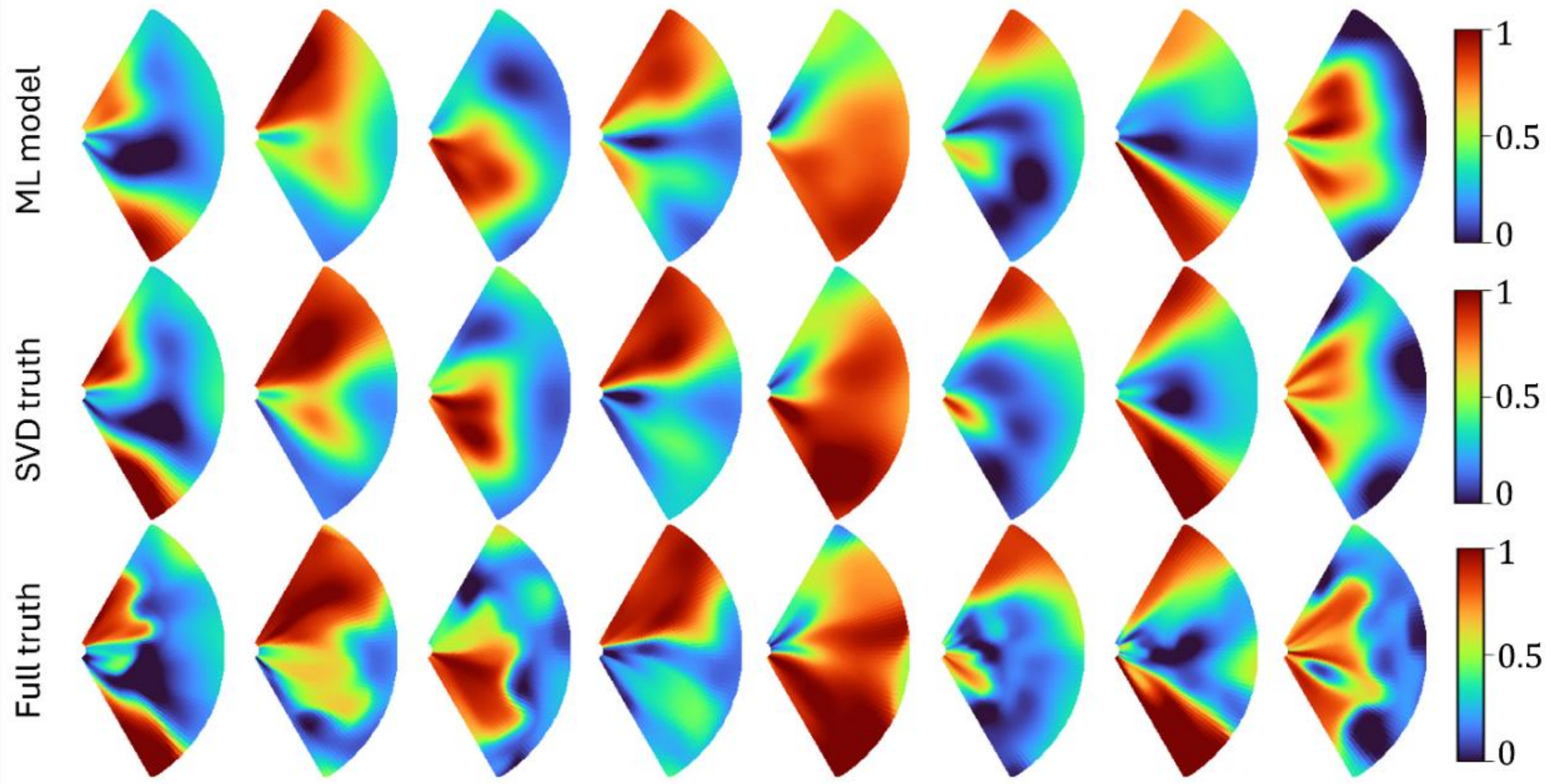

*Figure 3: ML-model reconstruction of $v_{mer}$ for randomly selected test snapshots. The ML prediction is compared with the rank-truncated SVD reconstruction using r = 30 retained modes (SVD truth) and the full WSA-ENLIL field (full truth).*

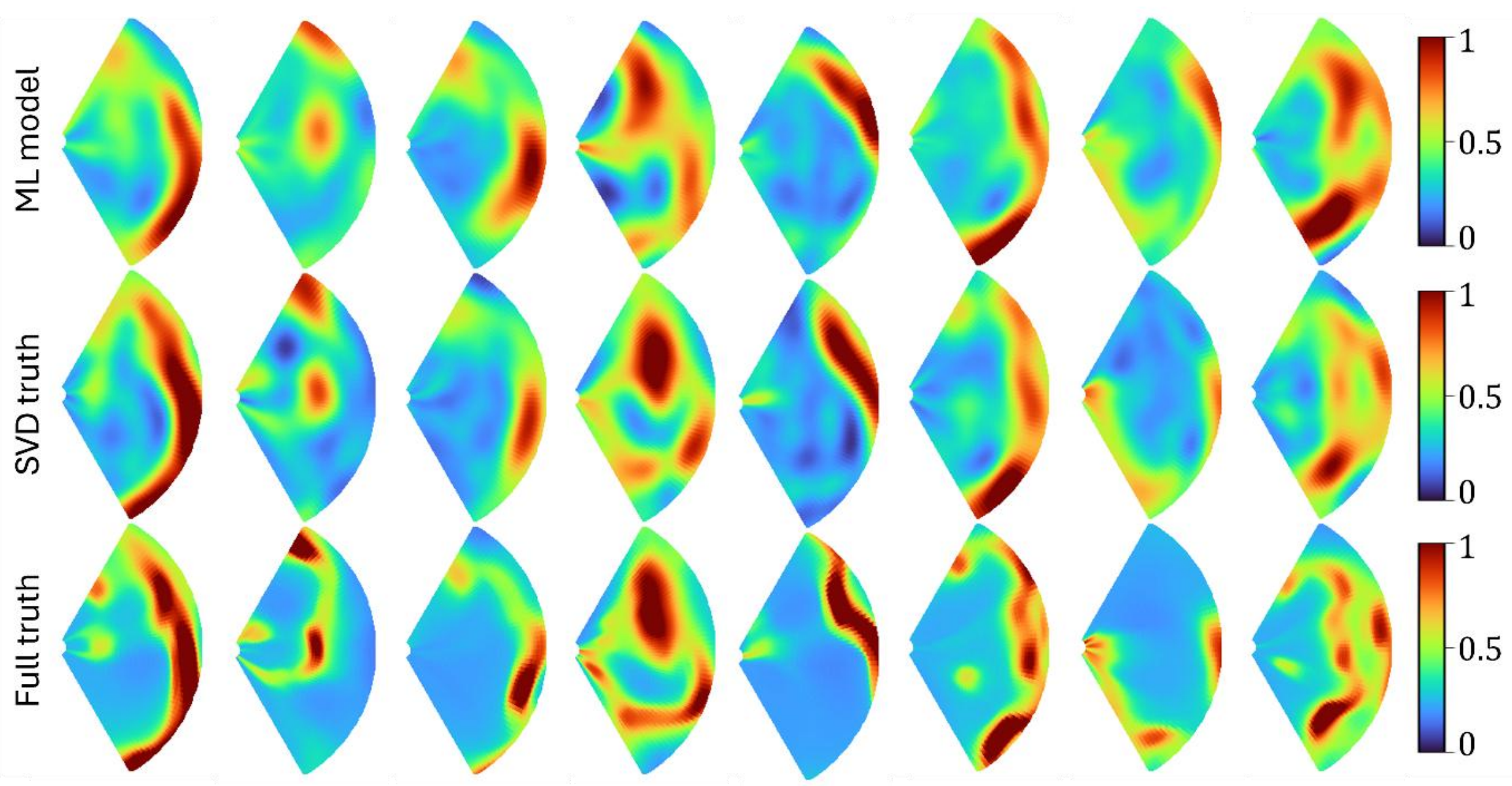

*Figure 4: ML-model reconstruction of $n_{mer}$ for randomly selected test snapshots. The ML prediction is compared with the rank-truncated SVD reconstruction using r = 30 retained modes (SVD truth) and the full-rank WSA-ENLIL snapshots (full truth).*

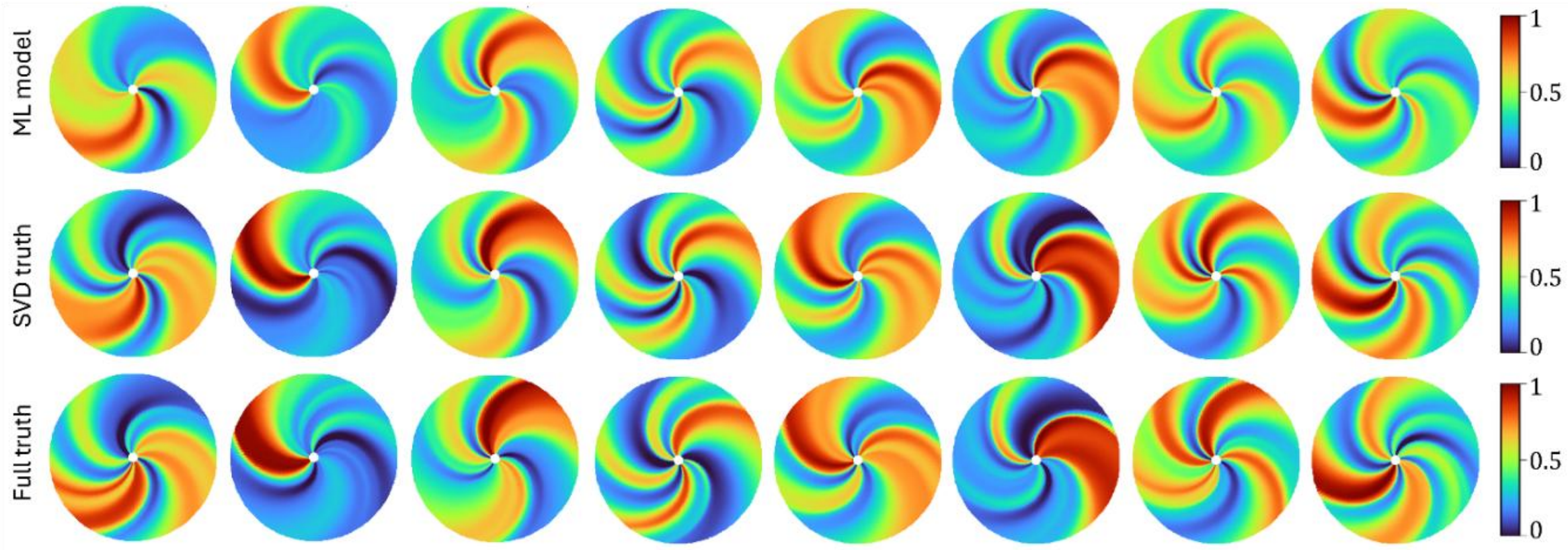

*Figure 5: ML-model reconstruction of $v_{eq}$ for randomly selected test snapshots. The ML prediction is compared with the rank-truncated SVD reconstruction using r = 30 retained modes (SVD truth) and the full-rank WSA-ENLIL snapshots (full truth).*

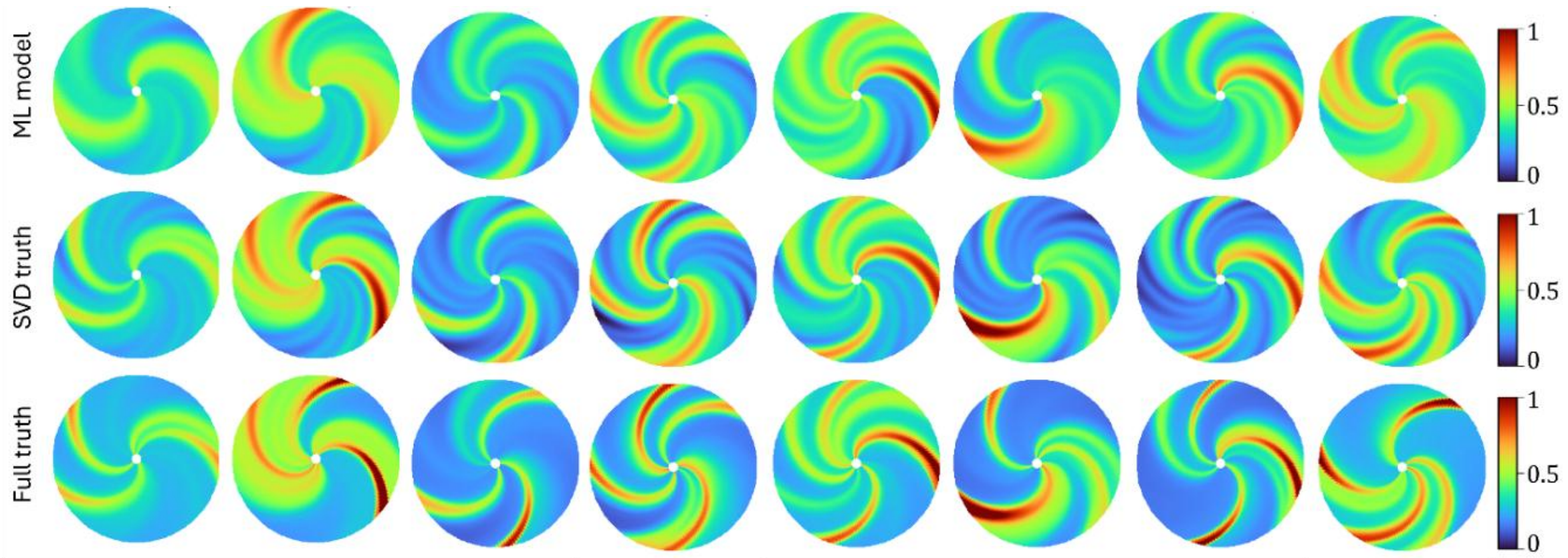

*Figure 6: ML-model reconstruction of $n_{eq}$ for randomly selected test snapshots. The model prediction is compared with the rank-truncated SVD reconstruction using r = 30 retained modes (SVD truth) and the full-rank WSA-ENLIL snapshots (full truth).*

To provide an additional assessment of the ML-model reconstruction performance, Figure 7 to Figure 10 compare the pointwise temporal variation of the reconstructed fields at six randomly selected spatial indices over the test interval. These locations were selected so as not to coincide with the positions of the virtual probes. The agreement with the rank-truncated SVD reference indicates that the model can capture the dominant features from sparse probe inputs. Deviations are more apparent during sharper transients and higher-frequency variations, where the finite retained rank and limited sensor information make the recovery of localized or rapidly varying features more challenging.

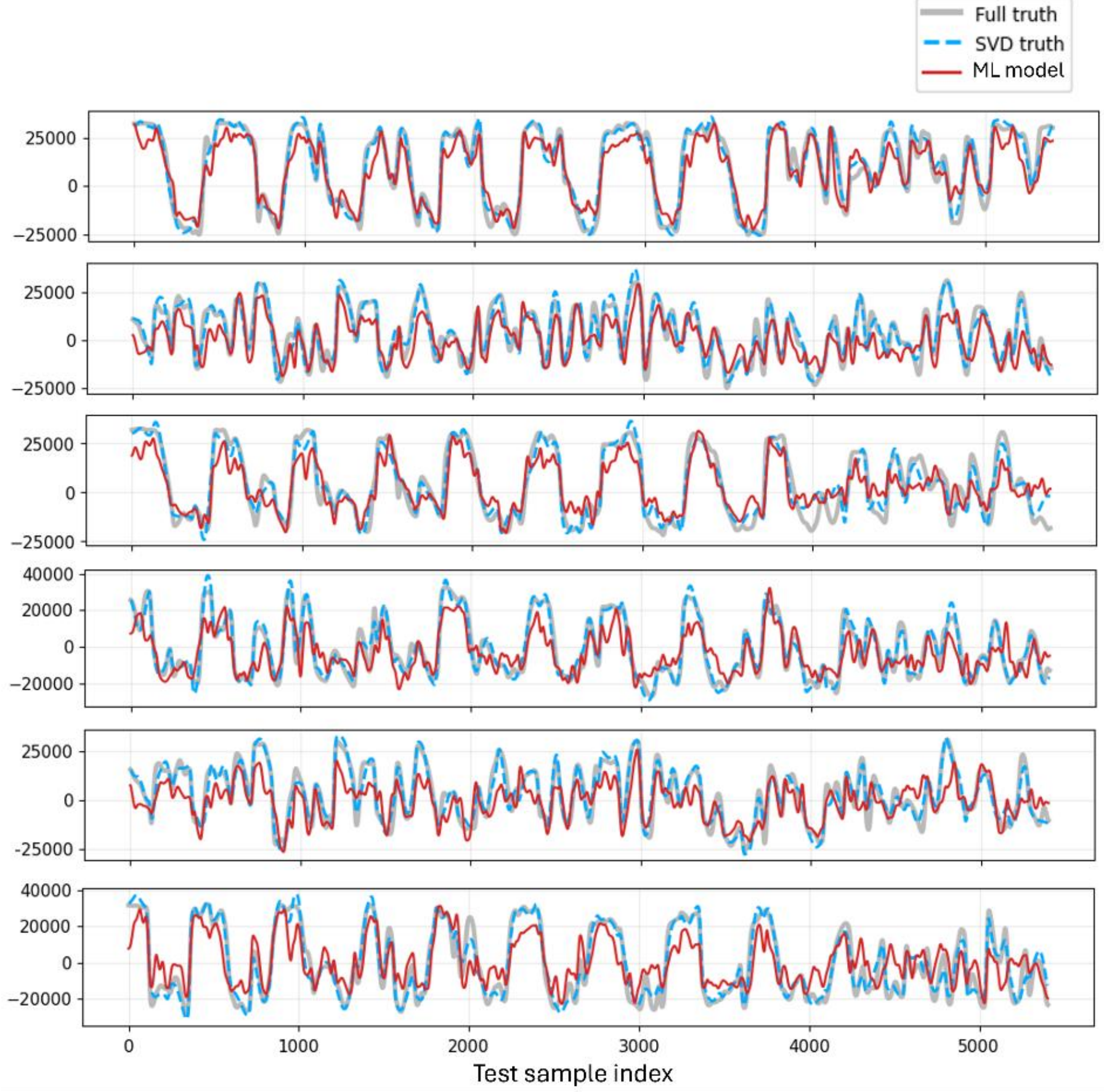

*Figure 7: Temporal variation of $v_{mer}$ obtained from the ML model at 6 randomly selected spatial indices over the test interval. The model prediction is compared with the rank-truncated SVD signal and the full-ranked WSA-ENLIL signal.*

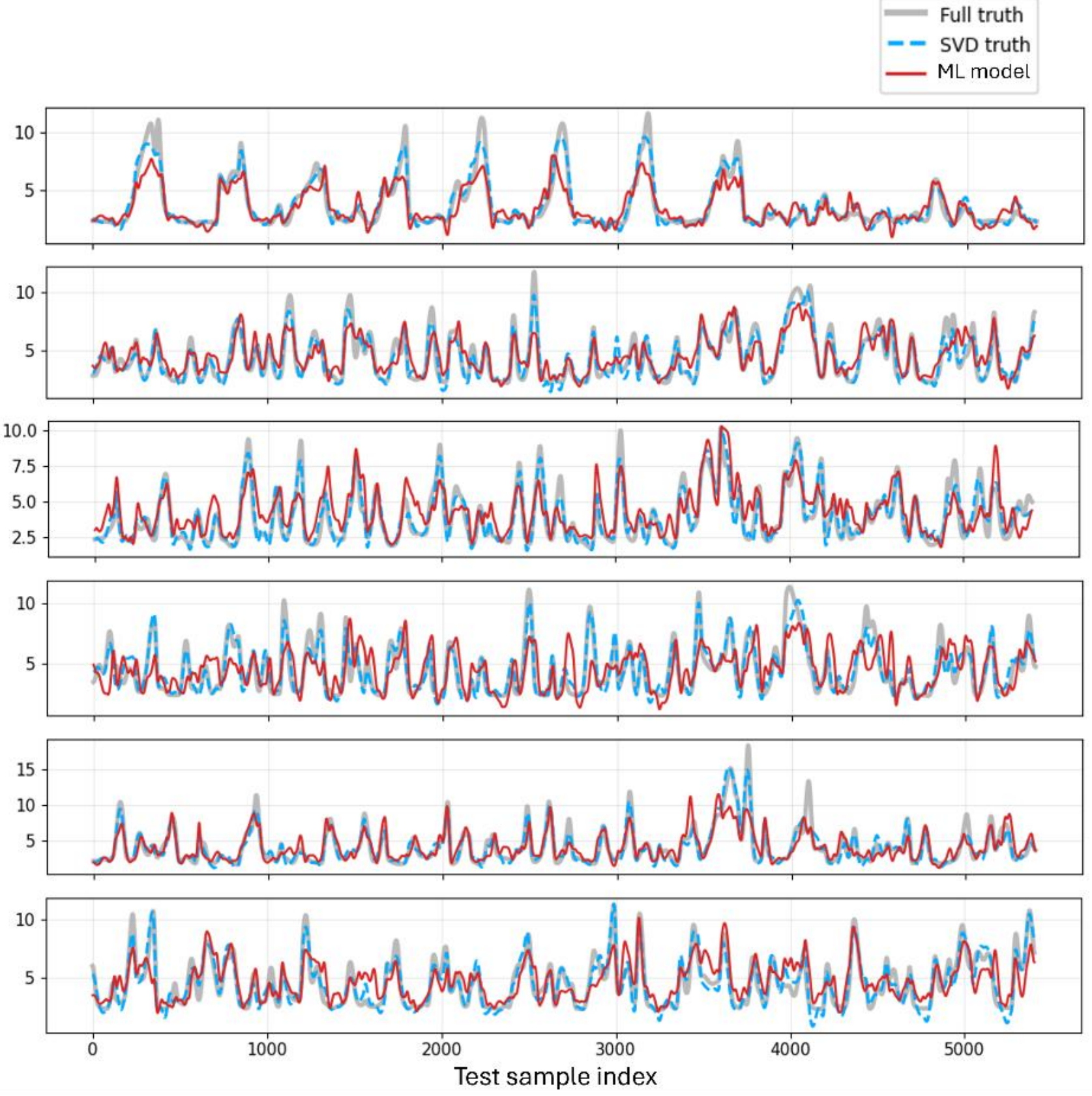

*Figure 8: Temporal variation of $n_{mer}$ obtained from the ML model at 6 randomly selected spatial indices over the test interval. The model prediction is compared with the rank-truncated SVD signal and the full-ranked WSA-ENLIL signal.*

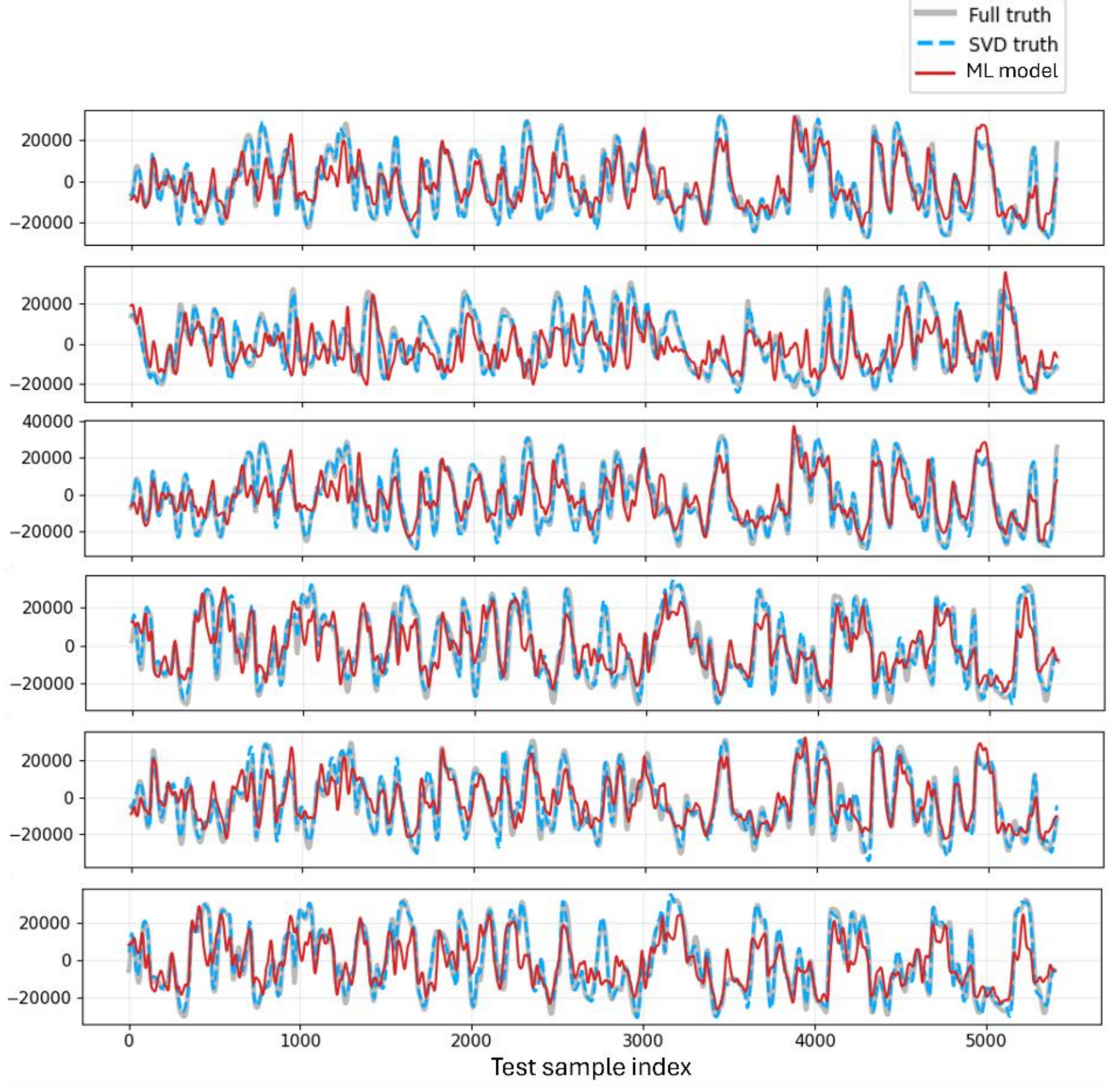

*Figure 9: Temporal variation of $v_{eq}$ obtained from the ML model at 6 randomly selected spatial indices over the test interval. The ML prediction is compared with the rank-truncated SVD signal and the full-ranked WSA-ENLIL signal.*

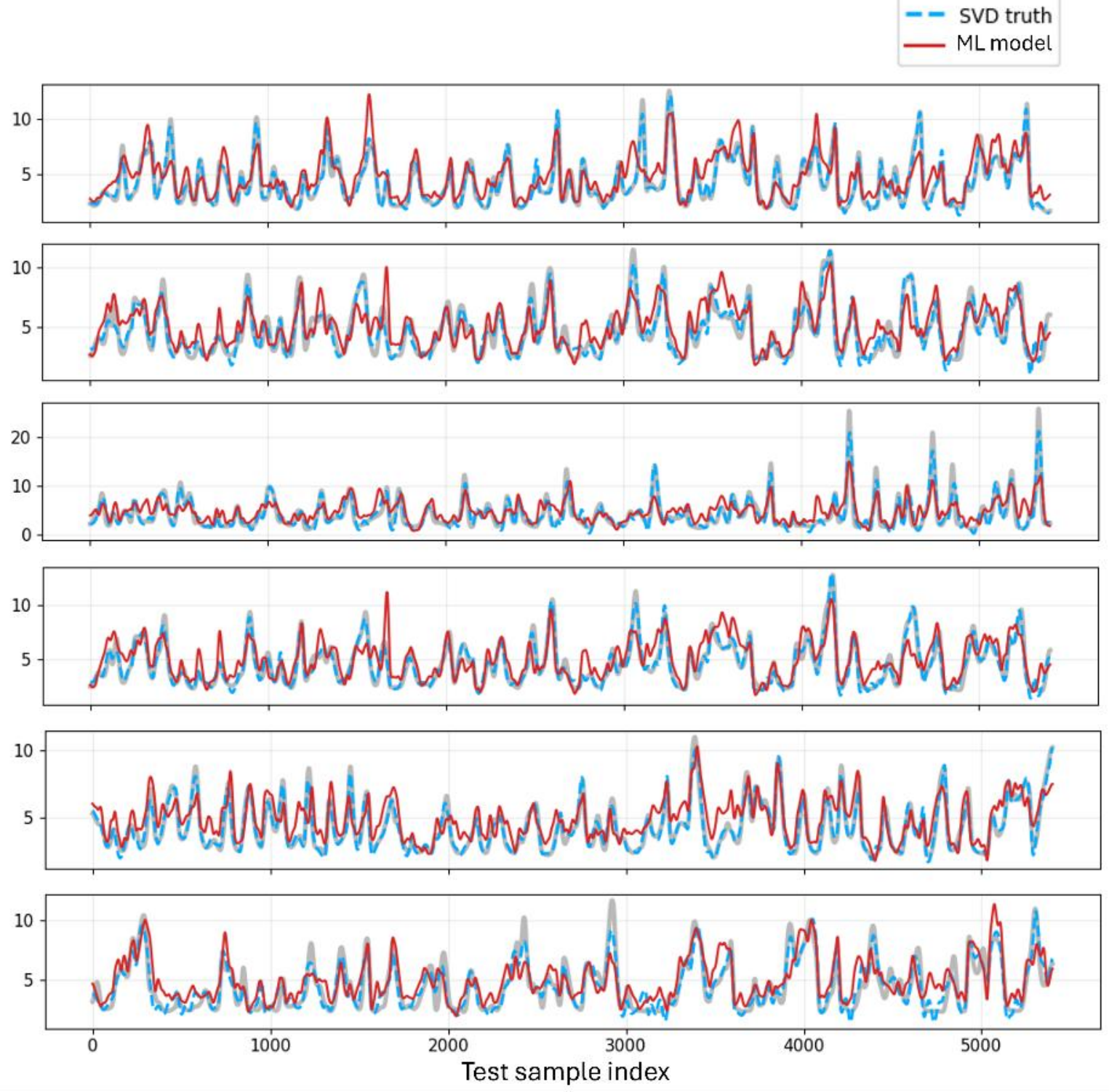

*Figure 10: Temporal variation of $n_{eq}$ obtained from the ML model at 6 randomly selected spatial indices over the test interval. The ML prediction is compared with the rank-truncated SVD signal and the full-ranked WSA-ENLIL signal.*

## 5. Model sensitivity studies

The effects of the ML model's main parameter choices were assessed through three separate sensitivity studies: (1) the number of retained SVD modes; (2) the number of input probe signals; and (3) the input signal time-lag length.

In each study, one parameter was varied while the other parameters were kept at their reference values. These reference values were 30 retained SVD modes, five input probe signals, and a lag length of 50 time steps. The same LSTM-decoder architecture and training settings described in Section 3.2 were used throughout.

For each parameter value and each target field, the model was trained five times using independently selected random sensor configurations and random initializations. The reconstruction error was then reported as the mean over the five repeated trainings, with the uncertainty shown as $\pm 1$ standard deviation. Errors were evaluated with respect to both the full-rank WSA-ENLIL fields and the corresponding rank-truncated SVD fields, allowing the total reconstruction error relative to the original data and the ML prediction error relative to the reduced representation to be distinguished.

### 5.1. Sensitivity to the number of retained SVD rank

The sensitivity to the reduced data representation was assessed by varying the number of retained SVD modes $r$ across the range of 3, 5, 10, 20, 30, 50, 70, 100 for the four plasma quantities to assess how this parameter affects the final reconstruction accuracy. This study examines the trade-off between the accuracy of the reduced SVD representation and the ML-model accuracy in inferring the increasingly smaller-scale features from sparse probe histories.

Figure 11 shows a clear trade-off in the sense that while increasing the rank improves the representational capacity of the SVD basis, it does not monotonically improve the ML-model reconstruction performance. At very low ranks, the reduced basis is too restrictive, and the SVD-truncated fields miss part of the spatial variability present in the full data. However, the ML model is more effectively able to infer these larger-scale features from input probe signals. As the rank increases, more spatial detail is retained, but the additional modes are generally associated with lower-amplitude, more localized, or faster-varying structures that are more difficult to infer from sparse probe histories. This can limit the ML model's reconstruction accuracy at higher ranks.

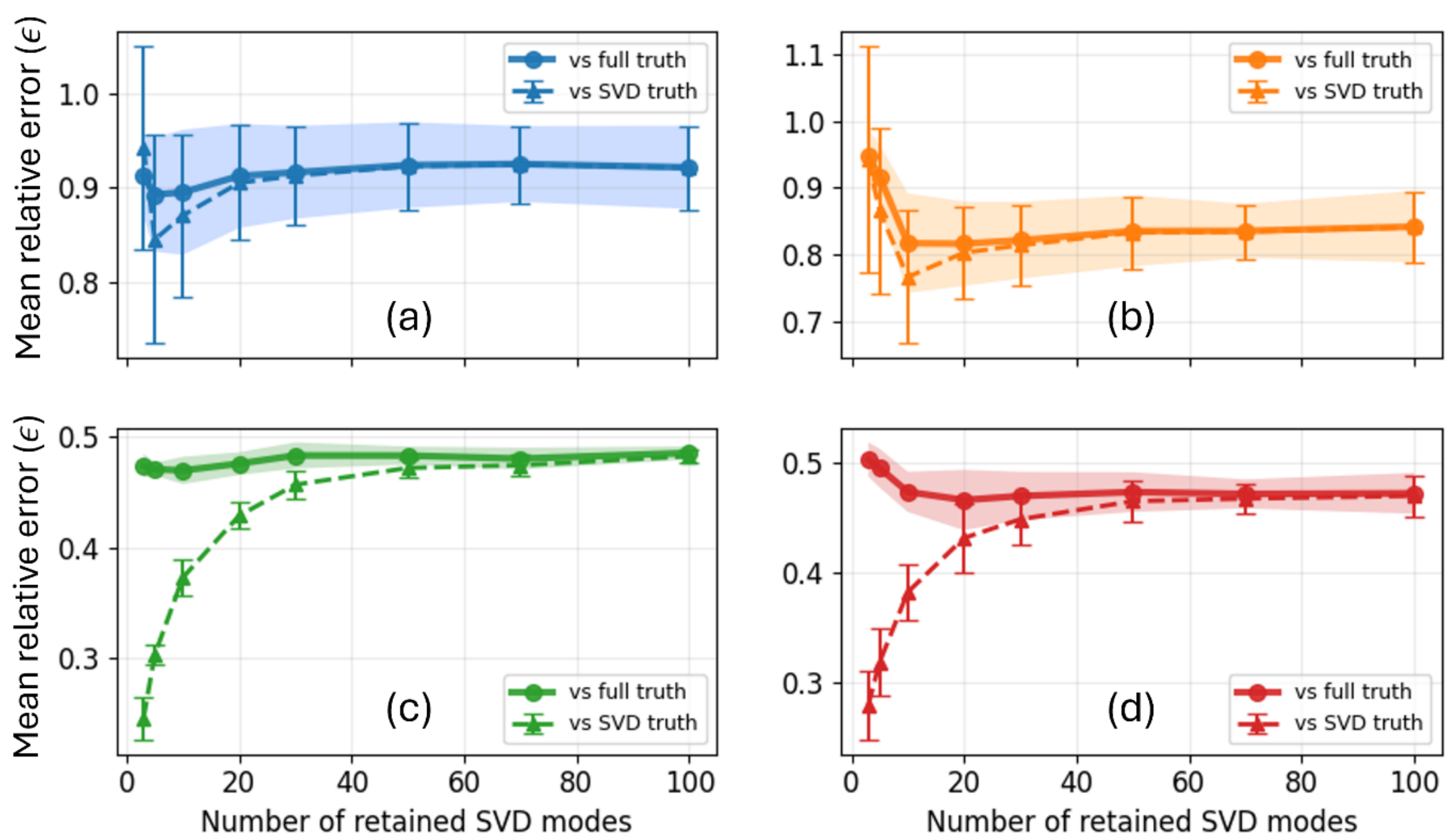

*Figure 11: Sensitivity of ML-model reconstruction error to the number of retained SVD modes. Mean relative error is shown for (a) $v_{mer}$, (b) $v_{eq}$, (c) $n_{mer}$, and (d) $n_{eq}$, evaluated with respect to both rank-truncated SVD fields and the full-rank WSA-ENLIL fields. Markers show the mean over five repeated trainings; error bars/bands indicate $\pm$1 standard deviation.*

Figure 12 provides a pointwise illustration of the same behavior at representative mid-domain locations. A higher number of modes introduces additional variations into the SVD reference, and thus the rank-truncated SVD traces more closely follow the full-rank WSA–ENLIL traces. However, these additional features are not always recovered more accurately by the ML model.

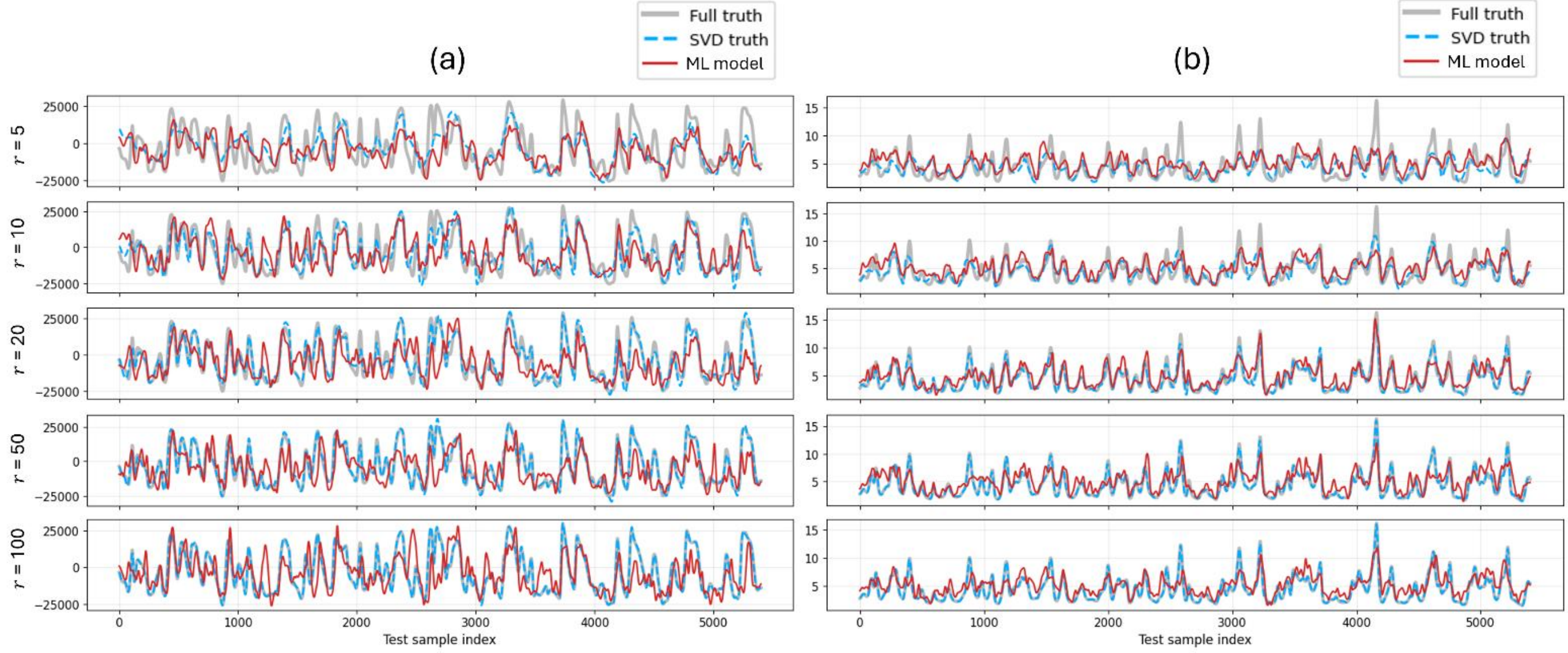

*Figure 12: Temporal variation of reconstructed fields at mid-domain location (as a sample) obtained from the ML model with different retained SVD ranks (k); (a) $v_{mer}$ at spatial index 15,616 and (b) $n_{eq}$ at spatial index 46,336 of the corresponding flattened snapshot vectors. ML predictions are compared with the associated rank-truncated SVD signals and full-rank WSA-ENLIL signals.*

Based on the behavior observed from this analysis, $r$ = 30 was selected for the main results presented in Section 4. This truncation rank value provides a compact reduced representation that retains the dominant spatial variability of all four plasma fields, while excluding the additional higher-rank modal coefficients which are otherwise to be captured by the ML model, providing a suitable representation-prediction accuracy trade-off.

### 5.2. Sensitivity to the number of sensors

The number of input probe signals was varied over the range of $m$ = 2, 3, 5, 10, 20, 30, 50, 70, 100 to examine the effect of sparse spatial sampling on the reconstruction accuracy. Figure 13 shows a consistent reduction in mean relative error for all four plasma fields as $m$ increases. The improvement is most pronounced when increasing from very sparse number of inputs to moderate sensor numbers, indicating that it is in the low-sensor regime where additional probe signals provide substantial new information.

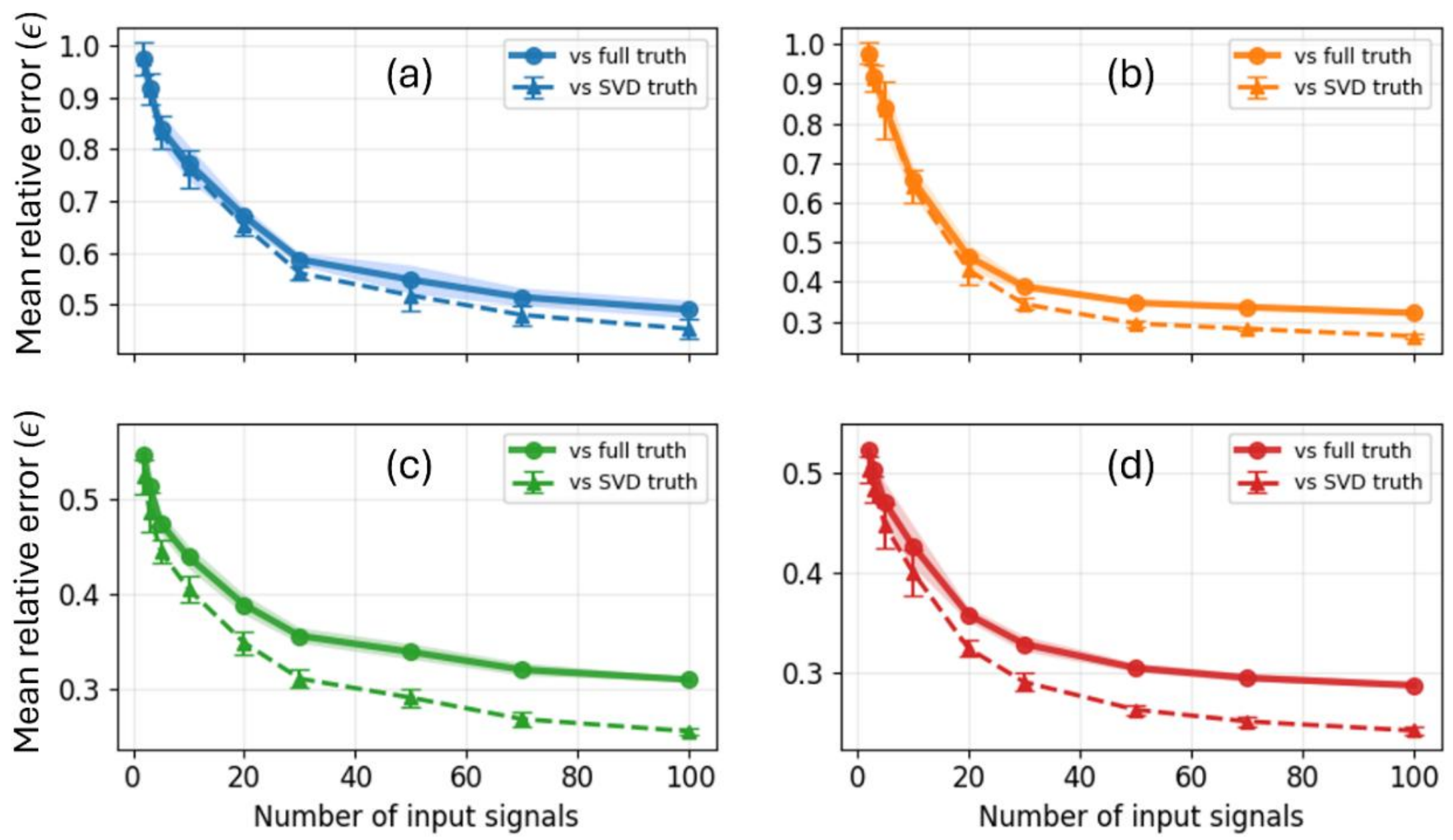

*Figure 13: Sensitivity of ML-model reconstruction error to the number of input probe signals ($m$). Mean relative error is shown for (a) $v_{mer}$, (b) $v_{eq}$, (c) $n_{mer}$, and (d) $n_{eq}$, evaluated with respect to both rank-truncated SVD fields and the full-rank WSA-ENLIL fields. Markers show the mean over five repeated trainings; error bars/bands indicate $\pm 1$ standard deviation.*

The error reduction becomes more gradual at larger $m$, particularly beyond approximately $m$ = 30, which suggests that the dominant recoverable information has already been captured, and additional probes provide diminishing returns in terms of reconstruction performance. Figure 14 further demonstrates this point by presenting the time variations of the meridional velocity and equatorial density as examples of the four fields at the representative mid-domain location. It is observed how increasing $m$ improves the agreement of the predicted amplitude and timing of sharper variations with the ground-truth traces.

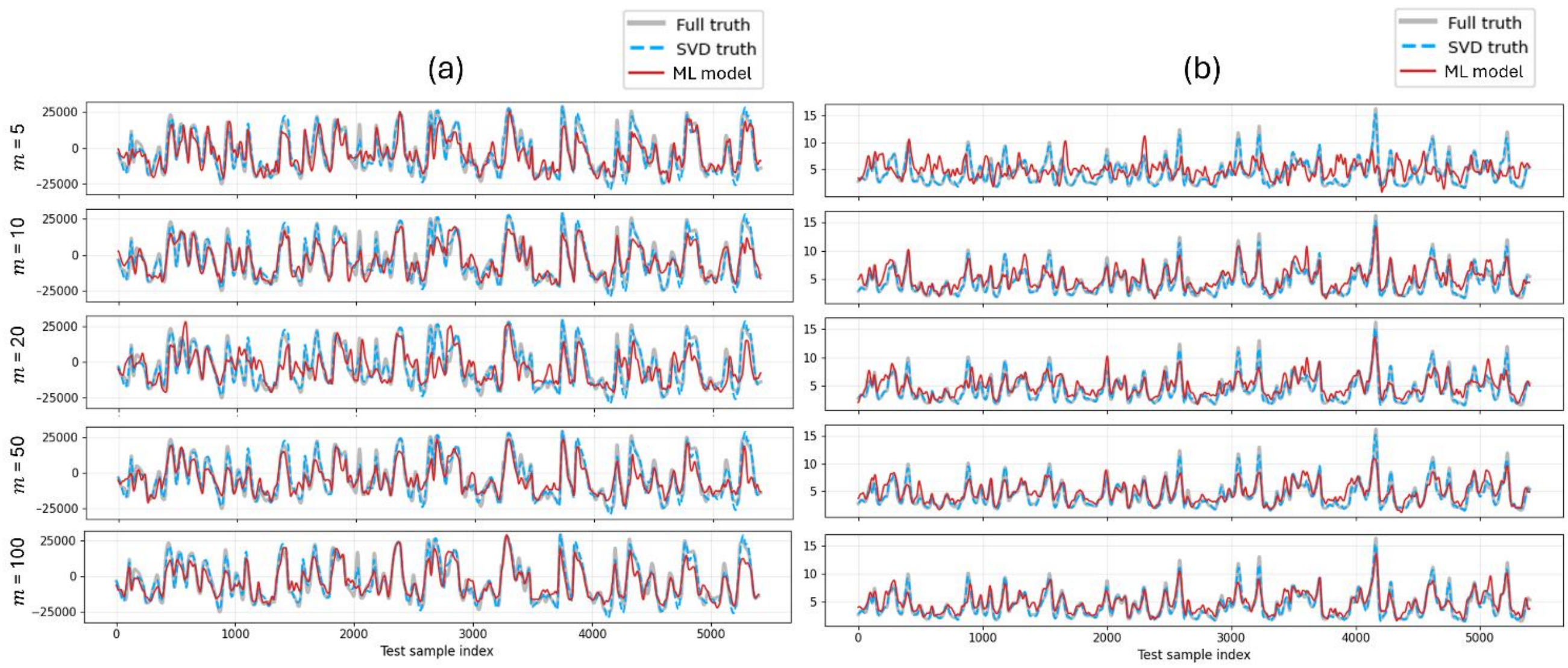

*Figure 14: Temporal variation of reconstructed fields at mid-domain location obtained from the ML model with different numbers of input sensor signals ($m$); (a) $v_{mer}$ and (b) $n_{eq}$. ML predictions are compared with the associated rank-truncated SVD signals and full-rank WSA-ENLIL signals.*

The main results in Section 4 were obtained using $m$ = 5 input probe signals, which intentionally represents a relatively strong sparse sensing configuration. Although higher sensor counts improve accuracy, the results show that useful full-field reconstructions can still be obtained from only a small number of input signals, which is the regime most relevant to sparse spacecraft probe measurements.

The variations represented by bars and bands across repeated trainings in Figure 13 reflect the sensitivity of reconstruction performance to sensor arrangement, since each repeated run used a different random set of probe location selections. This variability is most often the largest when only a few input signals are used, where the reconstruction depends strongly on whether the few selected probe placements provide a dynamically representative and sufficiently informative sampling of the domain. As the number of probes increases, the standard deviation generally decreases, indicating that the reconstruction becomes less sensitive to the sensor placement and more robust to random probe arrangement.

### 5.3. Sensitivity to input signal time-lag length

The ML model instances were retained over various input time-lag lengths over a range of $K$ = 5, 10, 20, 30, 50, 70, 100 to evaluate how much temporal history is required for reconstruction and how this parameter affects the model accuracy. Figure 15 shows that very short input histories generally lead to larger reconstruction errors, indicating that insufficient temporal context reduces the model's ability to infer the full reduced plasma state from sparse probe signals.

As the length of input signal increases, the error decreases up to an intermediate range around $K$ = 50-70 depending on the plasma field. This reflects the benefit of including enough history to capture the involved correlated time scales and their links to the plasma structures. However, increasing the lag length beyond this range leads to decorrelation and diminishes the accuracy of reconstruction slightly. This again suggests the existence of trade-off: the input window must be long enough to contain relevant correlated dynamics, but excessively long histories may include weakly correlated or decorrelated information that obscure the learning of the mapping between temporal signals and spatial structures during training.

Figure 16 presents this trend in terms of time variations at the representative mid-domain location. Short lag lengths overall show larger deviations from the ground-truth signals, while intermediate lag lengths give closer agreement in the amplitude and timing of the main variations. At the longest lag lengths, the improvement saturates or slightly degrades, consistent with the error trends seen in Figure 15. Based on this behavior, the main results in Section 4 were obtained using $K$ = 50 as a suitable balance between sufficient temporal context and avoiding unnecessarily long input histories.

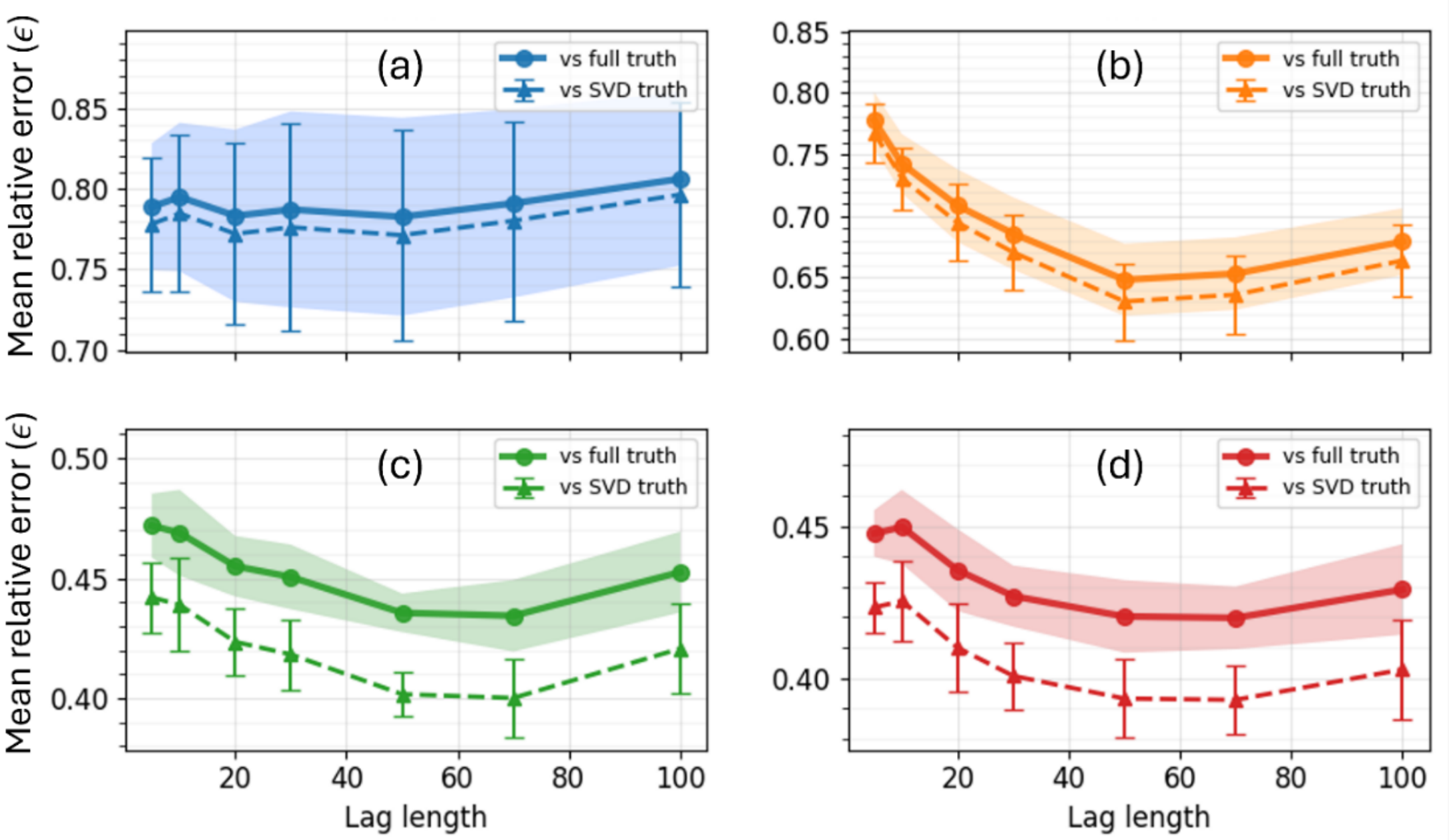

*Figure 15: Sensitivity of ML-model reconstruction error to the input time-lag length ($K$). Mean relative error is shown for (a) $v_{mer}$, (b) $v_{eq}$, (c) $n_{mer}$, and (d) $n_{eq}$, evaluated with respect to both rank-truncated SVD fields and the full-rank WSA-ENLIL fields. Markers show the mean over five repeated trainings; error bars/bands indicate $\pm$1 standard deviation.*

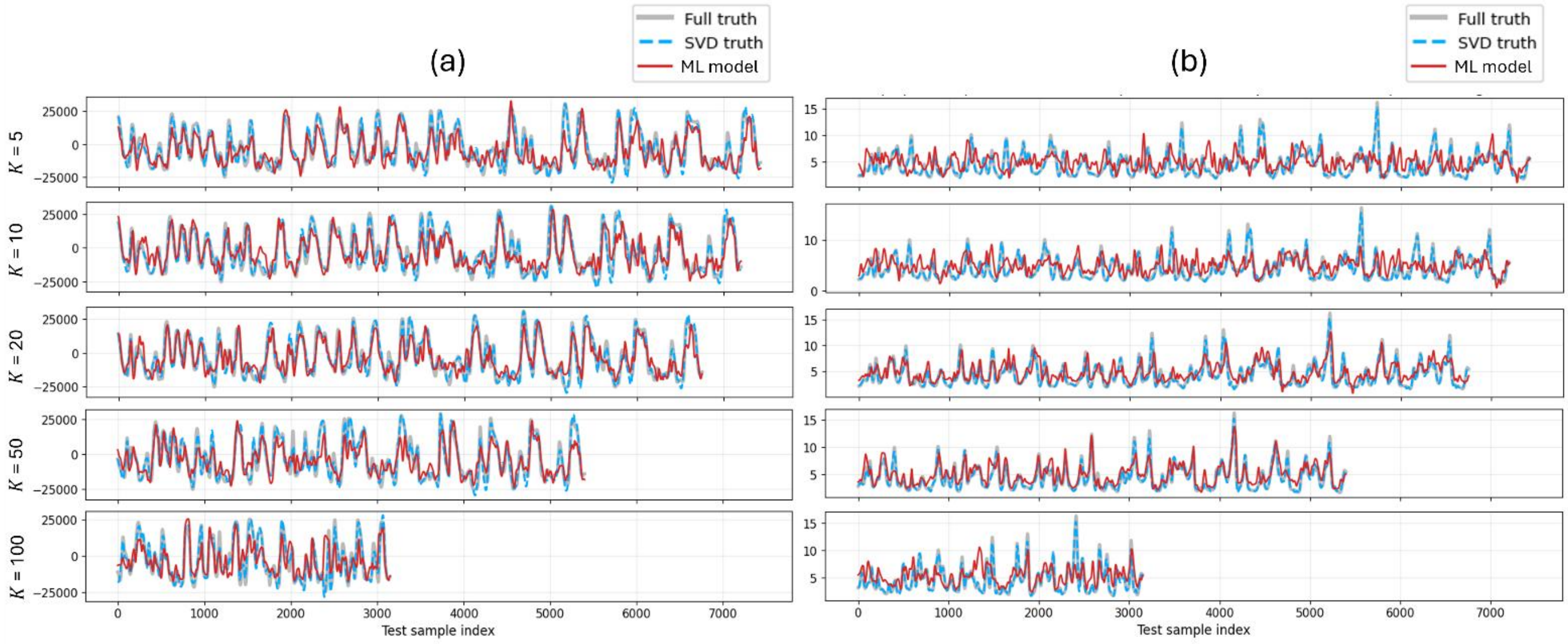

*Figure 16: Temporal variation of reconstructed fields at representative mid-domain location obtained from the ML model with different input time-lag lengths ($K$); (a) $v_{mer}$ and (b) $n_{eq}$. ML predictions are compared with the associated rank-truncated SVD signals and full-rank WSA-ENLIL signals. Note that the shorter traces at larger lag lengths arise because time-lagged samples were constructed within individual 169-frame WSA-ENLIL runs only; therefore, the number of valid windows per run decreases as (169- $K$+1), reducing the available test-sequence length for larger $K$.*

## Conclusions

In this work, we investigated the promise of ML-enabled compressed sensing for addressing the challenge of inferring spatially distributed solar-wind plasma structures from limited temporal probe measurements. To this end, a recurrent reduced-order learning framework was demonstrated using WSA-ENLIL solar-wind simulation data as a controlled test case, with the aim of assessing the utility of the method for reconstructing two-dimensional meridional and equatorial distributions of radial velocity and plasma density from a small number of virtual probe signals. The method was shown to recover the dominant radial and latitudinal variations in the meridional plane, the spiral-shaped solar-wind plasma organization in the equatorial plane, and spatial distributions of dynamically coupled plasma quantities not directly sensed.

The main methodological takeaway from this effort is that sparse temporal measurements can provide sufficient information for reconstructing the dominant large-scale organization of heliospheric plasma fields, but the accuracy of this reconstruction is governed by reduced-representation, sampling, and temporal-context trade-offs. The SVD truncation rank controls the balance between retaining spatial detail in the reduced-dimensional representation and preserving inferability from sparse measurements. Low-rank representations capture the dominant coherent structures and patterns but omit smaller-amplitude variability, while higher ranks retain more spatial detail but correspond to modes that are more localized, faster varying, and harder for the ML model to infer from limited probe histories. The number of probes similarly determines the amount of spatial information available to the model, with accuracy improving as more probe signals are added up to a limit beyond which higher sensor counts yield diminishing performance benefits. The input-history length also plays a non-monotonic role as short histories provide insufficient temporal context whereas excessively long histories may introduce decorrelated information that weakens the learned relation between local probe signals and distributed plasma structure. Among these observations, the reduced-representation trade-off points to an important direction for future work: how to refine the inference problem definition and correspondingly adapt the learning architecture so that smaller-amplitude and faster dynamical contributions, often associated with higher SVD ranks, can be captured more effectively without sacrificing robustness in the sparse-reconstruction regime.

Beyond the present methodological demonstration, the ability to extract high-dimensional spatial plasma-state information from limited probe measurements has direct relevance to plasma physics and space-weather science. Access to reconstructed velocity and density fields can support studies of how solar-wind structures form, evolve, and interact as they propagate through the heliosphere; in extended multi-plasma-field formulations, analogous reconstruction of pressure and magnetic-field distributions would further strengthen the connection to near-Earth plasma dynamics and magnetospheric driving. More broadly, reconstruction methods of the kind presented and demonstrated in this work can help bridge the gap between physics-based heliospheric simulations and sparse spacecraft observations, enabling local measurements to be used more directly in the analysis of high-dimensional plasma dynamics, turbulence, and solar-wind/magnetospheric interaction physics.

**Data Availability Statement**

The data that supports the findings of this study are available from the corresponding author upon reasonable request.